\documentstyle[12pt]{article}
\def\hybrid{\topmargin -20pt    \oddsidemargin 0pt
        \headheight 0pt \headsep 0pt
        \textwidth 6.25in       
        \textheight 9.5in       
        \marginparwidth .875in
        \parskip 5pt plus 1pt   \jot = 1.5ex}

\hybrid
\newskip\humongous \humongous=0pt plus 1000pt minus 1000pt
\def\caja{\mathsurround=0pt}
\def\eqalign#1{\,\vcenter{\openup1\jot \caja
        \ialign{\strut \hfil$\displaystyle{##}$&$
        \displaystyle{{}##}$\hfil\crcr#1\crcr}}\,}
\newif\ifdtup

\relax

\def\be{\begin{equation}}
\def\ee{\end{equation}}
\def\ba{\begin{eqnarray}}
\def\ea{\end{eqnarray}}

\begin{document}
\renewcommand{\theequation}{\thesection.\arabic{equation}}
\newcommand{\beq}{\begin{equation}}
\newcommand{\eeq}[1]{\label{#1}\end{equation}}
\newcommand{\ber}{\begin{eqnarray}}
\newcommand{\eer}[1]{\label{#1}\end{eqnarray}}
\begin{titlepage}
\begin{center}

\hfill CPTh-A323.0794\\
\hfill Crete-94-14\\
\hfill hep-ph/9409386\\
\hfill \\

\vskip .5in

{\large \bf  HIGGS-SECTOR SOLITONS }
\vskip .5in

{\bf C. Bachas} \footnotemark \\

\footnotetext{e-mail address: bachas@orphee.polytechnique.fr}

\vskip .1in

{\em Centre de Physique Th\'eorique\\
Ecole Polytechnique \\
 91128 Palaiseau, FRANCE}

\vskip .15in

       and

\vskip .15in

{\bf T.N. Tomaras} \footnote{e-mail address:
tomaras@plato.iesl.forth.gr }\\
\vskip
 .1in

{\em  Physics Department, University of Crete\\
and Research Center of Crete\\
 714 09 Heraklion, GREECE
  }\\

\vskip .1in

\end{center}

\vskip .4in

\begin{center} {\bf ABSTRACT }
\end{center}
\begin{quotation}\noindent
 We establish the existence of static, classically-stable,
   winding solitons
in   renormalizable  three-dimensional gauge models,
 with topologically trivial target space and  vacuum manifold.
 They are prototypes for possible analogous particle-like excitations in
the
higgs sector of the standard electroweak theory.
 \end{quotation}
\vskip1.0cm
September 1994\\
\end{titlepage}
\vfill
\eject

\setcounter{equation}{0}

Solitons  are perhaps the most spectacular
non-perturbative feature of field theories \cite{Rebbi}.
 Their presence is often
guaranteed by a conserved topological number to which one can
associate some minimum mass or energy. Although the
  standard electroweak theory has no such topologically-stable
solitons, it could still posses
classically-stable excitations that are sufficiently long-lived to be
relevant for cosmology.
The recent discussions of electroweak
vortex strings \cite{Vach} \cite{Perivol} underscore indeed
our ignorance of  the non-topological excitations
that can arise in gauge models.
In this letter we present some evidence
for the existence of non-topological
\footnote{Unlike   Q-balls \cite{Lee}
the non-topological solitons discussed here are static and uncharged
\cite{Raj} and owe
their stability to the dynamical exclusion of some region of configuration
space
 \cite{Mexican}.}
 particle-like solitons in
an extended higgs sector.

Most earlier discussions of electroweak solitons
 \cite{Tze} \cite{Fahri} \cite{Rub} \cite{Stern}
assumed   a  strongly-interacting higgs sector, whose would-be
goldstone bosons are described at low energy by an effective
non-linear $\sigma$-model. It has been argued that the corresponding
non-renormalizable lagrangian can have
stable solutions, which are
 identified with the technibaryons of an underlying
technicolor model,
 and whose distinguishing characteristic is the
non-trivial wrapping of the target SU(2) manifold \cite{Skyrme}.
This, of course, is
 at best a phenomenological description, since the
properties of the soliton cannot
be calculated reliably in a semiclassical expansion. The question that
arises naturally is whether such winding excitations can be
classically stable even for a {\it weakly-coupled} higgs sector.
In order to guide our thinking let us distinguish three
sources of potential instability:
\begin{itemize} \item{the winding can be undone if the higgs field passes
through
zero,}  \item{the evolution of the gauge fields can take us to a
winding-vacuum state plus radiation
\cite{thooft} \cite{Fahri} ,} and
\item{scalar-field excitations  loose their
 energy by shrinking to zero size \cite{Derrick}.}
\end{itemize}
The first (higgs) instability, which is not  considered in the
non-linear $\sigma$-model limit, imposes a lower bound on the
physical-scalar
mass times the would-be soliton size:
$$ m_H \rho > C \ . \eqno(1)$$
This follows by requiring the loss in potential energy $\sim \lambda v^4
\rho^3$
to exceed the gain in gradient energy $\sim v^2\rho$ when trying
to make the higgs vanish in the interior of the soliton.
The second (gauge) instability puts on the other hand an upper bound
on the gauge-boson mass times the would-be soliton size \cite{Rub}
\cite{Stern}:
$$ m_W \rho < C^\prime \ . \eqno(2)$$
This follows by requiring the loss in weak-magnetic energy $\sim 1/g^2\rho$
to exceed the gain in gradient energy when trying to turn on continuously
weak
gauge fields to reach the winding-vacuum state.
Here $v$ is the vacuum
expectation value of the higgs,  $\lambda$ its quartic self-coupling,
$g$ the gauge coupling, and $C$, $C^\prime$ numerical constants.
Taken together the above two bounds make a priori unlikely
the existence of winding
solitons in the perturbative minimal standard model.
 The prospects, however, are
better    in the presence of an extra higgs since the gauge instability,
and consequently the bound (2), are   in this
case absent. The validity of the above arguments was illustrated
explicitly in a toy two-dimensional model \cite{Mexican}.

 The issue which did not arise in this toy model is what can cure
the third (scale) instability and fix the soliton size.
Barring quantum effects \cite{Zee} \cite{Carlson} \cite{Tze},
 which would take us outside the semiclassical
treatment, a stabilizing role can only be played by the electro-weak
magnetic fields which are induced by the winding currents.
 There is an encouraging though not
conclusive argument that this may indeed happen in a two-higgs
standard  model.
It is  a paraphrasing of previous suggestions to stabilize the scale
of the soliton with extra heavy (hidden)
gauge bosons \cite{Dobado}. To simplify the argument let us freeze
  the magnitudes of the two doublets to some common
value $v$, and let $U$ be the relative $SU(2)$ phase.
We can obtain an energy functional for $U$
by solving the classical gauge-field equations in
a $\partial/m_W$ expansion.
The first two terms of this energy functional
are precisely those of the Skyrme model which has indeed stable winding
excitations
\cite{Skyrme}. Unfortunately  the argument is
  inconclusive because the size of the would-be soliton
turns out to be  $\rho\sim 1/m_W$,
thus invalidating our derivative expansion.
The issue  must thus be decided numerically in four dimensions
\cite{Tin}.
The purpose of this letter, on the other hand, is to demonstrate
analytically that gauge fields do stabilize winding solitons in an
analogous
three-dimensional model.

\hskip 0.6cm  Our starting point is the three-dimensional
$O(3)$ non-linear $\sigma$-model
 $$ S_0 = {v^2\over 2}\  \int   d^3x \
 \partial_\mu {\bf n} \cdot \partial^\mu {\bf n}  \eqno(3)$$
where ${\bf n}$ is a three-component scalar field subject to
the constraint
$$ {\bf n} \cdot {\bf n} = 1 . \eqno(4)$$
 We can solve the constraint    by
a stereographic projection of the three sphere onto the complex plane:
$$
 n_1+ i n_2 = {2\Omega \over 1 +
\vert\Omega\vert^2} \ \  ; \  \  n_3 = { 1 - \vert\Omega\vert^2 \over 1 +
\vert\Omega\vert^2} \ . \eqno(5)$$
 It is well known
that the above model  has static winding soliton solutions \cite{Belavin}
given by holomorphic functions $\Omega(z)$ where $z= x_1+ix_2$.
The solitons are classified by the number of times two-space wraps around
the target sphere:
$$ N = {1\over \pi} \int d^2x\  {  {\bar \partial} {\bar \Omega}
  \partial \Omega  -
 {\bar \partial}   \Omega   \partial  {\bar \Omega} \over (1 +
\vert\Omega\vert^2)^2 }  \ , \eqno(6)$$
where $\partial$ here stands for ${\partial\over \partial z}$.
The simplest solution,
$$\Omega^{sol} = {\rho e^{i\theta}\over z-z_0} + w_0 \ ,\eqno(7)$$
   describes a soliton with
  unit topological charge and energy $E^{sol} = 4\pi v^2$.
 It is characterized by six real parameters reflecting the invariance of
the underlying equations under the two-dimensional conformal group SL(2,C).
The complex parameter $w_0$ is in fact fixed by the choice of
boundary conditions at infinity: $w_0 = 0$ if ${\bf n}\rightarrow (0,0,1)$.
The remaining four collective coordinates correspond to translations,
U(1) rotations, and scale transformations of the soliton.

Let us next
   relax the non-linear constraint (4)
by introducing a mexican-hat potential.
  By Derrick's scaling argument \cite{Derrick}
  winding configurations are now unstable against shrinking
to zero size. Since we are interested in {\it renormalizable}
 models, we are
not allowed   to stabilize the size of the soliton  with
 explicit higher-derivative terms in the action
  \cite{Zak}.
   We must  thus try to evade Derrick's argument
by introducing gauge interactions. The simplest possibility is to
gauge  a  $U(1)$
subgroup of the global $O(3)$ symmetry of the model.
The corresponding gauge field can furthermore be massive
 without violating renormalizability,
provided it couples to a conserved current. We are thus led
to consider the following
  action:
$$\eqalign{
 S = \int   d^3x\  \Biggl[ {1\over 2}
 \mid (\partial_\mu + ie A_\mu) (\Phi_1+i\Phi_2)&\mid^2  +
 {1\over 2} \partial^\mu
\Phi_3 \partial_\mu \Phi_3   \cr
- & V(\Phi)  - {1\over 4}
 {\cal F}_{\mu\nu}^2 + { m^2\over 2} A^\mu
A_\mu  \Biggr] \ \ , \cr} \eqno(8)$$
with
$$ V(\Phi) =  {\lambda\over 4} (\sum_{a} \Phi_a\Phi_a - v^2)^2 +
{\kappa^2\over 8}
(\Phi_3-v)^{4} \ . \eqno(9)$$
Our choice of scalar potential   is
not the most general one consistent with the symmetry of the model, but
was dictated by later convenience.
Likewise, the mass of
the gauge field could arise from its
  coupling   with an extra complex scalar, but such a complication
will not be necessary for our purpose.
The  model defined by eqs. (8) and (9)
 has trivial topology, both in its scalar manifold and
in its vacuum sector. It reduces however to the ungauged
$O(3)$ non-linear $\sigma$-model
  in the naive
$$\lambda\rightarrow \infty \ \ {\rm and} \ \ e, \kappa\rightarrow 0
\eqno(10)$$
limit.
Our strategy will therefore be to show that
  for some range of parameters it has classically-stable solitons, which
 are small deformations
of the configuration (7) with $w_0 = 0$ and fixed size.

To this end, let us
   decompose the scalar triplet field
 into a radial and
an angular part: $\Phi_a = F n_a$, with ${\bf n}$   a  vector of unit
length
which can be expressed through $\Omega$ as in eq. (5).
Working in units of the   gauge-boson mass, $m=1$,
and rescaling:
 $F \rightarrow F  /\sqrt{2\lambda}$ and
 $ A_\mu \rightarrow  A_\mu  /\sqrt{2\lambda}$ ,
  brings  the
action to the form
$$
\eqalign{ S = {1\over 2\lambda} \int
  d^3x \ \Biggl[ {1\over 2} (\partial_\mu F)^2 &
+
 {1\over 2}F^2
  \vert (\partial_\mu + i {\tilde e} A_\mu)(n_1+i n_2)\vert^2
+ {1\over 2}F^2  (\partial_\mu n_3)^2
  \cr
 & - {1\over 4} (F^2 - m_H^2)^2
 -  {{\tilde\kappa}^2\over 8} (F n_3
-m_H)^{4} - {1\over 4} {\cal F}_{\mu\nu}^2 +
{1\over 2}  A^\mu A_\mu  \Biggr]
\ \ ,\cr} \eqno(11)
$$
with
$$   m_H \equiv \sqrt{2\lambda} v  \ , \ \
{\tilde e} \equiv e/\sqrt{2\lambda} \ \ {\rm and}
\ \  {\tilde\kappa} \equiv \kappa/\sqrt{2\lambda} \
 \  .
\   \eqno(12)$$
The above rewriting demonstrates that
  ${\tilde \kappa}, {\tilde e}$ and $m_H$ are the
only classically-relevant parameters of the model. The quartic
scalar coupling $\lambda$ on the other hand
 plays the role of Planck's constant
$\hbar$, and can be taken to zero
 independently in order to approach a semi-classical
limit. The
  existence of classically-stable winding solitons will not
therefore  be tied to the presence of
a strongly-interacting scalar sector.

To look for static minima of the energy we will proceed  in two steps:
we first keep the angular degree of freedom ${\bf n}$
 fixed  and  time-independent, and
minimize the energy with respect
to the radial and gauge fields $F$ and $A_\mu$.
 Assuming these stay close to their
vacuum values one finds:
$$ F  \simeq  m_H \Biggl[ 1  - {1\over 2 m_H^2}
  \partial_i {\bf n} \cdot \partial_i {\bf n}  \Biggr]  \ ,\eqno(13)
$$
$$A_0=0 \ ,\ \ {\rm and}\ \ \
A_k(x) \simeq 2{\tilde e} m_H^2 \int d^2y \ G_{kl}(x-y)
 J_l(y)  \   \ ,\eqno(14)$$
where
$$ J_l = {1\over 2}(n_2\partial_l n_1 - n_1\partial_l n_2) \eqno(15)$$
is the   U(1) current of the scalars, and
$$ G_{kl}(x) = \int {d^2p\over (2\pi)^2} \ e^{-i{\vec p}{\vec x}}\
{\delta_{kl}+p_kp_l\over {\vec p}^2+1}\  \eqno(16)$$
is the two-dimensional massive Green function.
Consistency of our approximation requires
 that
$$ {1\over m_H\rho}  \ll 1 \ ,\ \  {\tilde\kappa} m_H\rho \ll 1  \ , \ \
{\rm and}\ \  \ {\tilde e}m_H \ {\rm
min}(\rho,1) \ll 1 \ , \eqno(17)$$
with $\rho$ the typical scale over which ${\bf n}$ varies.
These conditions ensure in particular that $F-m_H \ll m_H$, and  that
${\tilde e} A_i {\bf n} \ll \partial_i {\bf n}$. They give
  a precise meaning to the naive limit, eq. (10).
Since $\rho$ will be determined   dynamically, we must
a posteriori check that these constraints can indeed be satisfied.

Eliminating $F$
 and $A_\mu$ with the help of eqs. (13) and (14)
 we arrive at an energy functional that depends only on the angular
degrees of freedom. It is of the form
$$ E =   E_0 - {\cal E} \eqno(18)$$
 where
$$ E_0 = {m_H^2\over 2\lambda}
\int d^2x \ {1\over 2} \partial_i{\bf n}\cdot \partial_i{\bf n}  \
\eqno(19)$$
is the energy in the non-linear $\sigma$-model limit, while
$$\eqalign{
 {\cal E} = {m_H^2\over 2\lambda}\
\Biggl[ \
  {1\over m_H^2}\
\int d^2x \  \Biggl( {1\over 2} &\partial_i{\bf n}\cdot \partial_i{\bf
n}\Biggr)^2
-{1\over 8}{\tilde\kappa}^2 m_H^2   \int d^2x\  (n_3-1)^4
  \cr
 & +  {\tilde e}^2 m_H^2   \int d^2x \int d^2y
J_i(x)G_{ik}(x-y)
J_k(y) \Biggr] \
,  \cr}
\eqno(20)$$
is   a   small perturbation under the above assumptions.
\vfil
\eject

Let us here pause for a minute and consider a simple calculus problem: we
are
asked to minimize a function of two variables
$G(u,v) = G_0(u,v) - {\cal G}(u,v)$,
 where $G_0$ has a line of degenerate
minima along the $u$ axis, while ${\cal G}$ is a small perturbation.
Minimizing first with respect to $v$ yields a line ${\bar v}(u)$
which lies a priori close to the $u$-axis. Along this line one finds easily
$$
 G(u,{\bar v}(u)) \simeq {\cal G}  -{1\over 2} {\cal G}^{\prime}
(G_0^{\prime\prime})^{-1}  {\cal G}^{\prime} + o({\cal G}^3) \ , \eqno(21)
$$
where the primes stand for derivatives with respect to $v$ and all the
functions on the right-hand side are evaluated at $v=0$.
As shown by this formula, for the expansion in powers of ${\cal G}$ to be
valid
   $G_0^{\prime\prime}$ must stay bounded away from zero, meaning that
the valley must not become too shallow in the transverse direction. In this
  case the first term of the series dominates, and   the
minima of the function $G$ are given by the minima of
  the perturbation ${\cal G}$ along the $u$ axis.

Going back  to the energy functional, eq.(18),  one notes that
the role of
$u$ is played by the  zero modes of  $\Omega^{cl}$, which is
  a local minimum of $E_0$, while the role of $v$ is played by the
infinite number of
 transverse fluctuations. Let us write
 ${\bf n} = {\bf n}^{cl}\sqrt{1-(\delta {\bf n})^2}+ \delta{\bf n}$
with ${\bf n}^{cl}\cdot\delta{\bf n}=0$, and consider fluctuations
which can be normalized on a sphere  of radius $\rho$, i.e. with respect to
the
inner product
$$ <\delta {\bf n} , \delta {\bf n}^\prime> \equiv
\int d\mu(x)\  \delta {\bf n} \cdot \delta {\bf n}^\prime \ \ ,
\ \ {\rm with} \ \ d\mu(x) \equiv {1\over\pi\rho^2}
 {d^2x\over
(1+\vert x\vert^2/\rho^2)^2}\ . \eqno(22)
$$
In the vicinity of ${\bf n}^{cl}$
the energy reads in an obvious notation:
$$ E-E^{cl} \simeq
-{\cal E}({\bf n}^{cl}) - \int d\mu \ {1\over 2}
 \delta{\bf n}^T \cdot E_0^{\prime\prime}
\cdot \delta{\bf n}
 - \int d\mu \  {\cal
E}^{\prime}\cdot
 \delta
{\bf n}  + o(\delta{\bf n}^3, {\cal E}\delta{\bf n}^2)  \ . \eqno(23)$$
 The matrix of quadratic
 fluctuations  $E_0^{\prime\prime}$,
has been shown in ref. \cite{spectrum}
 to have a discrete spectrum:
$\lambda^{(j,\alpha)} = j(j+1)-2$, where $j=1,2,...$ and $\alpha$
labels some finite degeneracy.
It is furthermore straightforward to check that
with the inner product (22) the first variation of the perturbation,
 ${\cal E}^{\prime}$, can
be normalized.
 The analysis of the calculus problem is under these conditions
 easily extended to show that we need only minimize the energy in the
space of zero modes of the unperturbed soliton, since
transverse fluctuations
affect the equations   at higher orders.

 Translation and rotation invariance ensures
in fact that the energy does not depend on the
U(1)-orientation and position. For any non-zero value
of ${\tilde\kappa}$ on the other
hand, the energy is infinite unless $w_0=0$.
The only relevant collective coordinate is
thus the scale, and   after a straightforward calculation we find
$$
 E(\rho)  = {2\pi m_H^2 \over\lambda}\
\Biggl[ 1 + {1\over 6} {\tilde\kappa}^2 m_H^2 \rho^2   - {4 \over
3 m_H^2 \rho^2}  -   {\tilde e}^2 m_H^2 \rho^2 \int_{0}^{\infty}dx\ {x^3
K_0^2(x) \over
x^2+\rho^2} \Biggr] \ , \eqno(24)
$$
 with $K_0$ the modified bessel function.
 The shape of the function $E(\rho)$,
up to overall multiplicative
 and additive factors,
 depends only on the two parameters
$$ a \equiv {{\tilde\kappa}^2 \over {\tilde e}^2 } \ \  {\rm and} \ \
 b\equiv {1\over {\tilde e}^2m_H^4} \ .
\eqno(25)$$
In the region above the thick   line of fig.1
$E$ grows monotonically with $\rho$ so that, to the extent that our
approximations are valid, we   conclude that
 the would-be soliton is unstable
against shrinking.  In the region below this thick line,
on the other hand, the function develops a local minimum at some
size ${\bar \rho}(a,b)$  at which the soliton is stabilized.
The tangents to the boundary of stability are
lines of constant ${\bar\rho}$ as shown  in the figure.
To complete
our proof of the existence of stable solitons, we must still
make sure that conditions (17) can be satisfied.
This can however always be arranged  by taking
  $m_H$ sufficiently large, while keeping $a$ and $b$ fixed at any point
below the thick line.
  Determining the complete region of
stability in the  $(m_H, {\tilde\kappa}, {\tilde e})$ space requires  a
numerical
investigation, which is  beyond the scope of the present letter.

Let us conclude  with a   comment on the potential importance of such
non-topological solitons, should they turn out to exist in the
electroweak model. Since they would decay by quantum tunneling, they
could be stable on cosmological time scales. Furthermore their expected
size is $\sim 1/m_W$,
their expected mass in the
$Tev$  range, while their annihilation cross-section, being essentially
geometrical, should be somewhat larger than weak cross-sections.
These properties would make them serious candidates for cold dark
matter in the universe.

\vskip 0.5cm

\vskip 0.3cm
{\bf Aknowledgements}
We thank L.Beljoudi, J.Iliopoulos, M.Karliner  and  L.Marleau
  for useful comments,   A.Sagnotti for a critical reading of the
manuscript, and
S. Trachanas for encouragement.   C.B. thanks
the Research Center of Crete  and the Physics Department of the
University of Crete, and T.N.T.
thanks the Centre de Physique
Th\'eorique of the Ecole Polytechnique
for hospitality while this work was being
completed. This research was supported in part by the EEC grants
SC1-0430-C and   CHRX-CT93-0340, as well as by the Greek General
Secretariat
of Research and Technology grant $91E\Delta 358$. .

\end{document}